\documentclass{IEEEcsmag}

\usepackage[hidelinks]{hyperref}
\expandafter\def\expandafter\UrlBreaks\expandafter{\UrlBreaks\do\/\do\*\do\-\do\~\do\'\do\"\do\-}
\usepackage{upmath,color}
\usepackage{booktabs}
\usepackage{arydshln}
\usepackage{xurl}
\hypersetup{breaklinks=true}
\usepackage{xcolor}
\newcommand{\review}[1]{\textcolor{black}{#1}}
\jvol{XX}
\jnum{XX}
\paper{8}
\jmonth{Month}
\jname{IEEE Software}
\jtitle{Feature Article}
\pubyear{2025}

\begin{document}

\sptitle{Feature Article}

\title{A Survey of Web Application Security Tutorials}

\author{Bhagya Chembakottu}
\affil{McGill University, Montreal, Canada}

\author{Martin P. Robillard}
\affil{McGill University, Montreal, Canada}

\markboth{FEATURE}{FEATURE}

\begin{abstract}
Developers rely on online tutorials to learn web application security, but tutorial quality varies. We reviewed 132 free security tutorials to examine topic coverage, authorship, and technical depth. Our analysis shows that most tutorials come from vendors and emphasize high-level explanations over concrete implementation guidance. Few tutorials provide complete runnable code examples or direct links to authoritative security resources such as the Open Web Application Security Project (OWASP), Common Weakness Enumeration (CWE), or Common Vulnerabilities and Exposures (CVE). We found that two visible signals help identify more useful tutorials: the presence of runnable code and direct links to official resources. These signals can help developers distinguish broad awareness material from tutorials that better support secure implementation.
\end{abstract}

\maketitle

\section{INTRODUCTION}

Web applications remain common targets of cyberattacks~\cite{cisa2025kev}. Attackers exploit injection flaws, weak session handling, and browser-side leaks to compromise these systems~\cite{Yao2023_WAE_Malware}. Developers can use Static Application Security Testing (SAST) tools to detect vulnerabilities, but these tools still miss some flaws and report false positives. Developers therefore need security expertise to interpret tool output and decide how to revise code~\cite{Li2023_ComparisonEvaluationSASTTools}. Vulnerabilities continue to recur year after year. This pattern shows that tools alone do not ensure secure software. Developers must also learn how to apply security knowledge during implementation~\cite{owasp2021top10}.

Developers learn secure web development from online tutorials and documentation~\cite{phptutorial}. These resources translate broad security principles into programming decisions. High-level principles such as input validation, least privilege, and secure session management offer conceptual direction, but developers still need concrete examples such as parameterized queries, restricted token scopes, or secure cookie settings. When tutorials provide clear and accurate examples, they help developers implement secure behavior. When tutorials omit details or present misleading guidance, they can normalize insecure patterns. Even minor oversights in example code can create security risks. For example, if a tutorial shows how to construct database queries without explaining how to handle user input safely, a developer may allow attackers to alter those queries and access sensitive data such as customer records. Likewise, tutorials that demonstrate authentication or token handling with hardcoded secrets can make credential forgery easier.

Software security therefore depends in part on what tutorials show and how clearly they explain implementation choices. Effective tutorials must do more than repeat advice such as ``validate user input.'' They must explain what that advice means in code. For example, a strong tutorial can show how an attacker enters input such as \verb|' OR 1=1--| to change the meaning of a database query. Because \texttt{1=1} always evaluates to true, the database may treat the condition as valid for every record and allow unauthorized access. To prevent this problem, tutorials should show concrete techniques such as parameterized queries or prepared statements, which force the database to treat user input as data rather than executable logic. Weak tutorials stop at vague statements such as ``add validation'' and do not explain how poor input handling leads to an exploit. That difference matters because it affects whether developers can translate a recommendation into safe code.

\review{
Security is communicated through different forms. Security professionals rely on established standards that define expected security properties and best practices, while developers learn from online tutorials that translate these ideas into code. However, a gap exists between these two sources. Standards are typically high-level and do not provide concrete implementation details, while tutorials vary widely in quality and often lack references to authoritative sources. As a result, developers may struggle to correctly interpret and apply security recommendations in practice. We therefore investigated how web security tutorials present security concepts and what practical signals they provide about their usefulness for implementation.
}

We analyzed the design and content of 132 web-application security tutorials. We compared each tutorial’s origin, structure, code examples, and references to authoritative security resources. We also grouped tutorials by authorship category (e.g., vendor-produced, independent blogs, and community-contributed sources) to examine whether the origin of a tutorial is associated with differences in e.g, code inclusion, citation practices, and presentation style. Our analysis highlights practical signals such as runnable code and direct links to official resources that can help readers identify tutorials that better support secure development.

\section{THE LANDSCAPE OF WEB APPLICATION SECURITY RESOURCES}

Security guidance for web developers falls into two broad categories. The first category includes official resources such as standards, taxonomies, and regulatory documents that define expected security properties. The second category includes web-accessible materials such as tutorials, blog posts, and framework documentation that explain how to implement those properties in specific technologies. This section describes both categories and explains why web applications remain vulnerable despite the large amount of available guidance.

\subsection{Official Security Resources}

Organizations rely on standards, best-practice documents, and vulnerability taxonomies to define and assess secure development. These resources provide a shared vocabulary for identifying, classifying, and prioritizing risks in web applications.

The \textit{Open Web Application Security Project (OWASP) Top 10} summarizes major categories of web application risk based on data from industry and research partners~\cite{owasp2021top10}. The list highlights recurring issues such as injection, broken authentication, and insecure design. \review{
The OWASP Top~10 classifies major categories of web application risks. It does not provide detailed implementation examples. However, the OWASP project recently added security cheat sheets with top 10 classes that include configuration advice and code examples for common vulnerabilities~\cite{owaspcheatsheets}.
}
The \textit{Common Weakness Enumeration (CWE)} catalog describes specific software weakness types. The \textit{Common Vulnerabilities and Exposures (CVE)} database assigns identifiers to publicly disclosed vulnerabilities. Public datasets such as the \textit{Known Exploited Vulnerabilities (KEV)} catalog, which the U.S. Cybersecurity and Infrastructure Security Agency maintains, connect CVE entries to real-world exploitation activity~\cite{cisa2025kev}.

Publications such as \textit{National Institute of Standards and Technology (NIST) Special Publication 800-53} define broader security and privacy controls for information systems~\cite{nist800-53}. \review{
Privacy regulations also influence web application design. or example, the General Data Protection Regulation (GDPR) defines requirements for handling personal data. Examples of sector specific regulations include the Payment Card Industry Data Security Standard (PCI DSS) and the Health Insurance Portability and Accountability Act (HIPAA).
}
These requirements directly affect how web applications handle personal, payment-related, and health-related data.

These documents offer broad coverage of risks and controls, but they target auditors, security architects, and compliance officers more than developers. Developers usually encounter them through tutorials, code snippets, or framework documentation that interpret these standards for programming tasks. We do not claim that this list is exhaustive. We selected these resources because researchers, industry reports, and empirical studies refer to them often in discussions of web security practice~\cite{phptutorial,guler2024_atropos}.

\subsection{Web-Accessible Security Tutorials}

In practice, many developers learn secure web development from web-accessible 
tutorials, framework guides, and community posts rather than from official standards. Web searches for phrases such as ``secure cookie handling in Django'' or ``authentication practices for Spring Boot'' return many pages that present instructional content. These pages appear on vendor websites, non-profit initiatives, learning platforms, and personal blogs. Tutorial depth and structure vary widely. Tutorials differ from standards documents because authors publish them without formal review processes. Authors may combine instruction with product promotion, and sites update content at different rates.

\review{
Developers may also use generative artificial intelligence systems together with tutorials and documentation. Prior research shows that developers encounter reliability problems, incomplete explanations, and difficulty validating generated output when they use such tools~\cite{llm-application}. Other studies show that secure code generation depends on prompting strategies and contextual guidance~\cite{prompting}. Prior work comparing large language model answers with Stack Overflow responses for Android permission-related questions reports differences in correctness and completeness of the information provided~\cite{llm-stackoverflow}. The study highlights that responses from the two sources may vary in the level of detail, use of code examples, and alignment with platform-specific practices, which can affect how easily developers can interpret and validate the solutions. Large language models may also depend on existing tutorials and documentation because these materials contribute to the data used during model training.
}

\subsection{Need for Developer-Centric Security Learning}

Official security resources provide detailed taxonomies, control catalogs, and regulatory requirements, but developers still face difficulty when they try to translate these high-level prescriptions into correct implementation decisions. Research on vulnerability detection shows that tool output alone does not ensure security. Developers must still understand how to revise code and configuration correctly~\cite{Yang2019_TowardsBetterUtilizingSAST,Li2023_ComparisonEvaluationSASTTools}. Tool combinations can improve vulnerability detection, but developer-facing guidance remains a limiting factor~\cite{AlKassar2023_WHIP}.

Web tutorials attempt to address this gap by offering concrete examples that developers can follow. However, prior work shows that insecure or incomplete examples can spread into real systems at scale~\cite{phptutorial}. When tutorials misuse security libraries or oversimplify configurations, they risk normalizing insecure coding practices.

In the remainder of this article, we examine web-accessible security tutorials, characterize their structure, use of code, and citation of authoritative sources, and evaluate how these characteristics may help developers find reliable security guidance.

\section{METHODOLOGY}
\label{sec:landscape}

\begin{figure}
\centerline{\includegraphics[width=18.5pc]{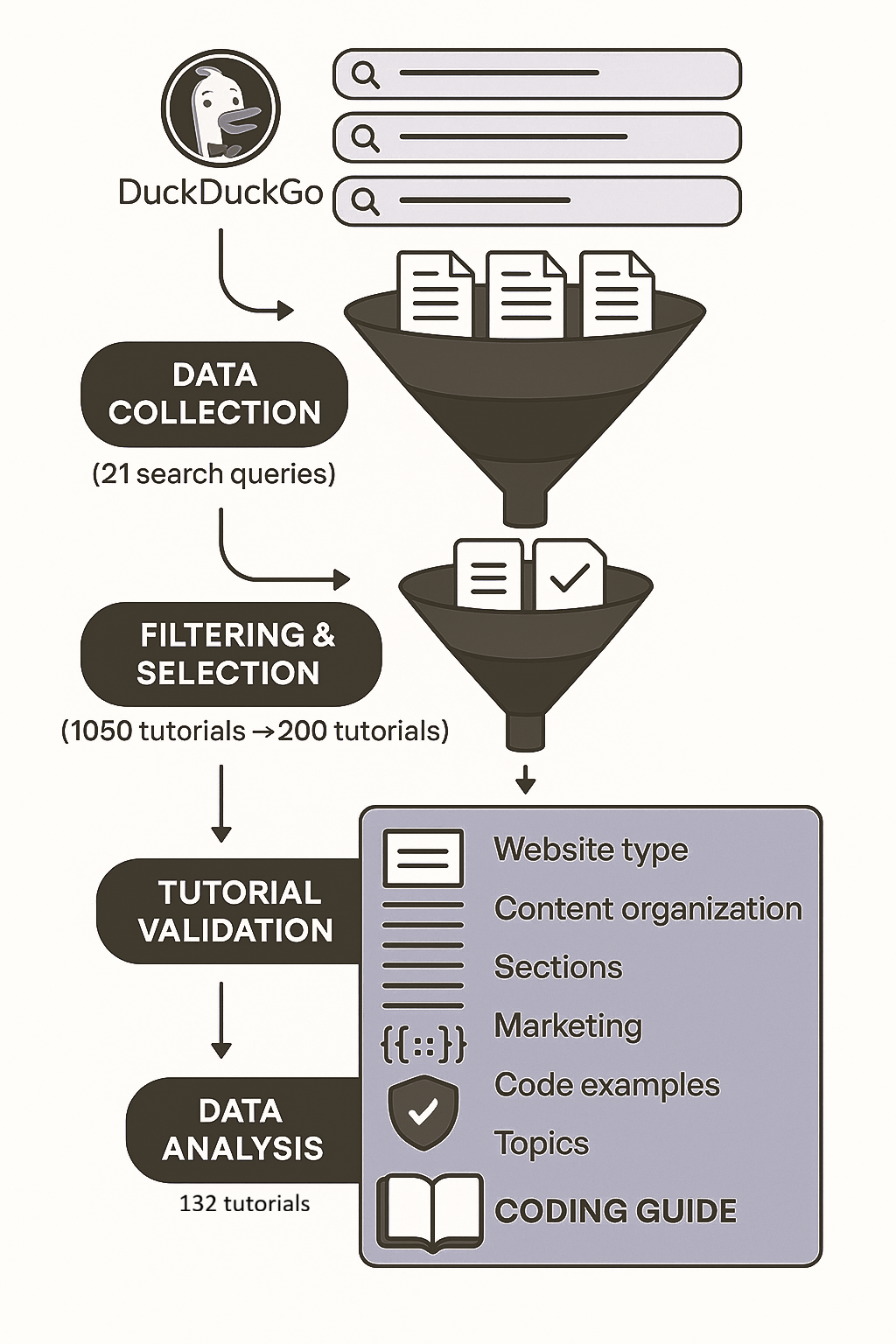}}
\caption{Study Design}
\label{f:pipeline}
\end{figure}

We identified web-application security tutorials through web search. We selected tutorials that appeared frequently and ranked highly across multiple queries. We then characterized the tutorials along several dimensions with a structured coding scheme. Figure~\ref{f:pipeline} summarizes the study design.

\subsection{Data Collection}

We collected tutorials through web search to approximate what developers encounter when they seek guidance on securing web applications.

\review{
We used the \texttt{duckduckgo-search} API to retrieve search results. Major search engines apply extensive personalization and ranking adjustments based on user history. These mechanisms make it difficult to reproduce search results across environments. DuckDuckGo reduces user tracking and limits personalization effects. This property allowed us to obtain results not directly biased by personalization algorithms.
}

\paragraph{Query construction.}
We tested two preliminary query sets to evaluate keyword diversity and saturation.
Set~1 included 30 broad phrases such as \textit{``app security,'' ``cross-site scripting prevention tutorial,'' ``web development security,''} and \textit{``XML attacks security tutorial.''}
Set~2 narrowed the set to 21 queries centered on the phrase \textit{``web application security''} and common variants such as \textit{``web application security checklist,'' ``how to secure a web application,''} and \textit{``backend security.''}
Across these trials, new queries added between 11 and 49 unique URLs per iteration. After the 20th query, the rate of new URLs dropped. We therefore adopted the 21-query configuration because it preserved topic diversity without adding substantial redundancy. Our research artifact includes the full query list.

\paragraph{Corpus retrieval and validation.}
For each of the 21 queries, we retrieved the first 50 results, which produced an initial corpus of 1{,}050 entries.
After removing duplicates, we obtained 872 distinct URLs.
To assess the reliability of the \texttt{duckduckgo-search} API, we compared its output with the DuckDuckGo web interface for three representative queries.
Across the top ten results per query, we found 23 overlapping results out of 30 total positions, which corresponds to a 77\% micro-averaged overlap.
We calculated this ratio by aggregating all positions across queries before we computed agreement so that the estimate reflected overall similarity rather than per-query variation.

\paragraph{Ranking and selection.}
To quantify visibility across different search intents, we ranked each tutorial URL by its position in every query result. For each of the 21 queries, the DuckDuckGo API returned URLs and we assigned a rank from 1 to 50 according to position, where 1 indicates the highest-ranked tutorial. If a URL did not appear for a query, we assigned rank 51 to represent non-appearance. When different subpages from the same domain appeared, such as \texttt{domain.com/a} and \texttt{domain.com/a/b}, we treated each subpage as a separate tutorial because the pages differed in structure, content, or detail and because search engines returned them independently.

We then computed a \textit{total visibility score} for each URL by summing its ranks across all queries.
Lower scores indicate higher overall visibility. Using the script in our research artifact, we sorted all URLs by aggregated visibility score and selected the top 200 most visible URLs. We then manually excluded videos, courses, PDF handbooks, marketing landing pages, non-instructional GitHub repositories, automated scanner outputs, and other non-tutorial formats. After we applied these criteria, 132 URLs remained in the final dataset.

\subsection{Data Analysis}

For each of the 132 tutorials, we extracted structured attributes that described origin, presentation style, and technical depth. These attributes include website type, content organization, presence of runnable or illustrative code examples, presence of advertisements, references to widely used security standards and taxonomies, and overall tutorial length. Table~\ref{tab:summary-merged} summarizes the variables. Our research artifact provides the coding guide and data collection procedure.

\review{
\paragraph{Coding Scheme and Inter-Rater Reliability.}
We developed a structured coding guide to ensure consistent characterization of tutorials. Both authors independently applied the coding guide to the same subset of ten tutorials. We measured agreement using raw percent agreement and Cohen’s~$\kappa$. Objective variables such as advertisement presence, code examples, content organization, and references to security standards showed almost perfect agreement. Disagreements occurred only for two variables: \textit{Tutorial Focus} and \textit{Tutorial Type}. These cases involved tutorials that combined multiple instructional styles or mixed vendor and community content.
}

\review{
\paragraph{Narrative Length Extraction.}
We measured narrative length using the visible instructional text of each tutorial. We removed HTML markup, scripts, navigation elements, and other structural components. We also removed code blocks so that the length measurement reflected explanatory prose only. We used whitespace tokenization to estimate word counts. 
}

\review{\paragraph{Statistical Analysis.}
We analyzed relationships between tutorial characteristics and narrative length, and associations among categorical attributes. Given most tutorials are short, but a few are very long, distribution of narrative length and the mixture of variable types, we used nonparametric statistical tests throughout. We compared narrative length across tutorial groups using the Mann--Whitney U test and examined associations between categorical variables using Pearson's chi-square test. We also report 95\% bootstrap confidence intervals for descriptive comparisons.} 

\review{We applied a Bonferroni correction and interpret statistical significance using a corrected threshold of 0.0125. All results in this article follow this criterion. The companion website and replication package provide the complete statistical outputs, visualizations, and analysis scripts used in this study~\cite{replication}.}

\section{FINDINGS}\label{sec:findings}

\subsection{Dataset Overview}

Our final dataset contains 132 web application security tutorials. Table~\ref{tab:summary-merged} summarizes the coded characteristics and reports counts and mean narrative length for each category.

Vendor-authored tutorials account for 73\% of the dataset, which reflects the strong visibility of commercial documentation in search results. Community, educational, and media sources account for the remaining 27\%. Most tutorials take the form of best-practice or overview articles (57\%), while 43\% take the form of how-to guides, reference pages, or short learning modules.

Most tutorials emphasize process-level guidance (66\%), such as secure workflows, configuration, and deployment practices, rather than code-level implementation detail. Code examples remain rare: 80\% contain no code blocks, 15\% include only a short snippet, and 5\% provide pseudocode or a runnable example. Advertisements appear in 82\% of tutorials through banners or promotional elements. OWASP is the most frequently referenced framework (54\%).
\subsection{Tutorial Length}

We examined how narrative text length varied across tutorial categories. We define \textit{narrative length} as tutorial text excluding code blocks and markup, so that the measure captures explanatory prose only. Table~\ref{tab:summary-merged} reports descriptive statistics for each category.

Across major categories, average narrative length differed only modestly. Tutorials that included code averaged 2.8k words, compared with 3.1k words for tutorials without code. Vendor tutorials averaged 2.8k words, while non-vendor tutorials averaged 3.7k words. Tutorials with advertisements averaged 2.8k words, whereas tutorials without advertisements averaged 3.9k words. \review{
Our statistical analysis did not indicate consistent differences in narrative length across these tutorial groups.
}

We treat these differences as descriptive rather than as direct measures of instructional depth. Narrative length measures how much text a tutorial contains, but it does not measure how well that text supports implementation. A short tutorial may provide a concise and useful example, while a longer tutorial may repeat high-level advice without adding implementation detail. We therefore use narrative length to contextualize other characteristics, such as code inclusion and references to authoritative sources, rather than as a quality measure on its own.

\textit{Tutorials of similar length can differ substantially in usefulness. Longer text does not guarantee clearer or more actionable guidance.}

\subsection{Code Examples}

Code examples were uncommon. Only 26 tutorials (20\%) contained any form of code, whether a short snippet, pseudocode, or a runnable example. Of these 26 tutorials, three provided runnable examples that readers could execute directly, and three presented pseudocode that illustrated control flow or logic. Tutorials with code showed only a small tendency toward greater narrative length.

\review{
Our statistical analysis did not indicate a clear relationship between code inclusion and narrative length.
}
Even so, code examples often demonstrated concrete practices such as parameterized queries, secure session configuration, or token validation. These examples supplied implementation detail that narrative explanation alone did not provide.

\textit{Code examples are uncommon in security tutorials. However, even small examples can help developers understand how to implement a recommendation in practice.}

\subsection{Commercial Framing}\label{sec:ads}

Commercial framing appeared throughout the dataset. Advertisements were present in 108 tutorials (82\%).

\review{
Vendor-authored tutorials showed a strong association with advertisement presence. We also observed a relationship between advertisement presence and code inclusion. Tutorials that contained code displayed advertisements less often than narrative-only tutorials, although both patterns appeared in the dataset.
}

Despite their prevalence, advertisements did not differentiate tutorials clearly in terms of narrative length or tutorial type. Instead, advertisements primarily signaled the commercial origin of the resource.

\textit{Advertisements were nearly ubiquitous in vendor tutorials, but their presence did not reliably indicate shorter or less detailed content. Developers may treat advertisements mainly as a signal of origin rather than of instructional depth.}

\subsection{Vendor and Non-vendor Tutorials}

Vendor tutorials dominated search results and therefore dominated the dataset. Vendor and non-vendor tutorials differed somewhat in content structure. Vendors more frequently include overview  or best-practice summaries, whereas non-vendors more frequently include how-to material or framework-specific guidance.

Non-vendor tutorials were descriptively longer on average (3.7k words vs.\ 2.8k words).
\review{
Our statistical analysis did not indicate a consistent difference in narrative length between vendor and non-vendor tutorials.
}

\textit{Vendor tutorials were more numerous, but they were not consistently shorter or less detailed than non-vendor tutorials.}

\begin{table*}[t]
\centering
\caption{Summary of tutorial characteristics and text length comparisons (\(n = 132\)). Mutually exclusive categories sum to 132; multi-label categories do not.}

\label{tab:summary-merged}
\small
\begin{tabular}{@{}lccc@{}}
\toprule
\textbf{Dimension / Category} &
\textbf{Count (\%)} &
\textbf{Mean Length (k words)} &
\textbf{Notes} \\
\midrule

\multicolumn{4}{l}{\textbf{Mutually Exclusive Tutorial Characteristics}} \\[2pt]

\multicolumn{4}{l}{\emph{Tutorial Source}} \\
\cdashline{1-4}[.4pt/2pt]
Vendor & 97 (73.3\%) & 2.8 & Dominant source type \\
Community / Educational / Media & 35 (26.7\%) & 3.7 & Minority share \\[4pt]

\multicolumn{4}{l}{\emph{Tutorial Structure}} \\
\cdashline{1-4}[.4pt/2pt]
Best-practice / Overview & 76 (57.3\%) & 2.7 & Most common format \\
How-to / Reference / Module & 56 (42.7\%) & 3.6 & Longer, detailed formats \\[4pt]

\multicolumn{4}{l}{\emph{Code Inclusion}} \\
\cdashline{1-4}[.4pt/2pt]
No code & 106 (80.3\%) & 3.1 & Narrative only \\
Short snippet & 20 (15.15\%) & 2.8 & Minimal example fragment \\
Runnable example & 3 (2.3\%) & 2.8 & Complete, executable code \\
Pseudocode & 3 (2.3\%) & 2.8 & Illustrative logic only \\[4pt]

\multicolumn{4}{l}{\emph{Commercial Framing}} \\
\cdashline{1-4}[.4pt/2pt]
Advertisements present & 108 (81.8\%) & 2.8 & Banner or call-to-action visible \\
Advertisements absent & 24 (18.2\%) & 3.9 & Non-commercial presentation \\[6pt]

\midrule

\multicolumn{4}{l}{\textbf{Multi-Label Tutorial Characteristics}} \\[2pt]

\multicolumn{4}{l}{\emph{Tutorial Focus}} \\
\cdashline{1-4}[.4pt/2pt]
Process-level topics & 88 (66.4\%) & 2.8 & Deployment, configuration, workflow \\
Tool- or code-level focus & 27 (20.6\%) & 2.8 & Implementation-oriented \\[4pt]

\multicolumn{4}{l}{\emph{Security Framework References}} \\
\cdashline{1-4}[.4pt/2pt]
OWASP (name only / linked) & 71 (54.0\%) & 3.0 & 33\% name-only, 21\% linked \\
Other frameworks (any) & 28 (21.0\%) & 3.6 & GDPR, PCI DSS, HIPAA, etc. \\
Authoritative sources & 27 (20.5\%) & 3.7 & Official domains (e.g., \texttt{owasp.org}) \\

\bottomrule
\end{tabular}
\end{table*}

\section{DISCUSSION}

We interpret these findings for developers and teams that rely on tutorials during secure development work.

\subsection{\review{What Developers Might See When They Search}}

Common search queries return a landscape dominated by vendor tutorials, with a smaller share of community and educational sources. Across these sources, many tutorials repeat similar high-level descriptions of risks and recommended practices. Developers who browse multiple search results therefore encounter repeated lists of vulnerabilities and repeated descriptions of configuration guidance.

Only a small subset of tutorials include code or link directly to primary security standards. These tutorials appear less often in search results but provide more concrete examples or references to authoritative sources. Our data therefore show a visible divide between broadly accessible awareness material and more implementation-oriented content.

\subsection{How Many Tutorials Are Worth Reading?}

This overlap raises a practical question for developers: how many tutorials must they read before they gain useful information? Many tutorials present similar narrative guidance with only small differences in organization or emphasis. After a developer reads one or two tutorials of this type, additional tutorials may add little new information.

Tutorials that include code or direct links to standards provide a different kind of value. Code examples show how to apply a recommendation in a specific language or framework. References to standards give developers a way to verify that a local recommendation aligns with broader classifications of risk or control. Because these features appear less often, they can help identify tutorials that add new information instead of restating familiar points. Developers who choose one or two tutorials with code or authoritative sources are more likely to encounter concrete implementation guidance than developers who scan several narrative-only tutorials.

\subsection{Simple Heuristics for Selecting Tutorials}

Our analysis suggests two practical heuristics that developers can apply when they scan search results. The first heuristic concerns \emph{the inclusion of code examples}, which our data captures through the categories \textit{no code}, \textit{short snippet}, \textit{pseudocode}, and \textit{runnable example}. Although code presence was not associated with longer text, tutorials with snippets or runnable examples gave clearer illustrations of how to implement a security recommendation. Even a small example gave developers a concrete starting point for experimentation or adaptation.

The second heuristic concerns \emph{references to authoritative security resources}. Our variables distinguish name-only mentions such as ``OWASP'' from direct links to official sources such as \texttt{owasp.org} or \texttt{cwe.mitre.org}. Tutorials with such links appeared less often in search results, but they tended to provide more context and clearer connections between specific advice and broader risk classifications.

These heuristics reflect visible differences in how tutorials present security guidance. Tutorials that include code examples or direct links to authoritative sources more often provide concrete artifacts such as configuration fragments, API usage patterns, or explicit references that developers can inspect, adapt, or cross-check. When many tutorials look similar at first glance, these two signals offer a simple way to prioritize pages that provide stronger support for implementation decisions.

Tutorial authors and organizations can draw a parallel lesson from these findings. They can improve tutorial usefulness by including concrete code or configuration examples and by linking those examples directly to primary security standards instead of relying only on high-level summaries.

\section{LIMITATIONS}

Our dataset captures a visible subset of English-language web security tutorials, but several methodological decisions limit the scope of the results.

\begin{itemize}
  \item \textbf{Scope of media.} We excluded videos, books, and long-form courses and focused on web-based textual tutorials only. Developers also learn security concepts from video platforms, interactive courses, and code repositories that provide runnable examples. Our dataset therefore represents one segment of the broader developer learning ecosystem.

  \item \textbf{Impact of the search engine.}
  \review{
  We collected tutorials through the DuckDuckGo search API to reduce personalization and support reproducible results. Search engines rank pages using many signals, including search engine optimization (SEO), commercial promotion, and automated content generation sometimes described as AI-assisted SEO. These mechanisms influence which tutorials appear in search results and may promote marketing-oriented content. Our dataset therefore reflects visibility for one search engine rather than the exact rankings that developers observe across different search engines, locations, or browsing contexts.
  }

  \item \textbf{Temporal scope.} The dataset reflects the state of the web at the time of collection. Authors update tutorials, publish new materials, and reorganize documentation over time.
  \review{
  Generative AI tools may increase the publication rate of tutorial-style content and automated documentation. This change could alter the composition of search results over time.
  }

  \item \textbf{Granularity of measures.} Our coding process measures whether tutorials contain code and what type of code they contain, but it does not evaluate code correctness or executability. Likewise, the ``authoritative sources'' variable captures provenance, such as \texttt{owasp.org}, but it does not assess whether authors interpret linked content accurately.
\end{itemize}

\section{CONCLUSION}

We surveyed 132 web application security tutorials to examine what developers are likely to encounter when they search for web application security guidance online. Our analysis shows that most tutorials are vendor-hosted, frequently include advertisements, and mainly present high-level descriptions of risks and recommended practices. Only a small subset provide concrete implementation guidance through code examples or direct links to authoritative security resources such as OWASP or CWE. Narrative length alone does not reliably indicate whether a tutorial offers practical implementation detail.

\review{
Developers who examine many search results need simple ways to identify useful tutorials. Our analysis shows that two signals help identify tutorials that provide implementation guidance: the presence of code examples and direct links to recognized security resources. These signals do not guarantee correctness, but they consistently appeared in tutorials that provided concrete implementation detail. Tutorial authors and organizations can improve instructional material by including code examples and linking recommendations to primary security standards. These additions help developers interpret, verify, and apply security guidance during implementation.
}

\section{ACKNOWLEDGMENTS}
We thank our colleagues for feedback on the study design and early drafts of this article. Jun Soo Kim helped pilot the study design. This work is funded by NSERC.

\bibliographystyle{IEEEtran} 
\bibliography{references} 

\begin{IEEEbiography}{\includegraphics[width=1.1in]{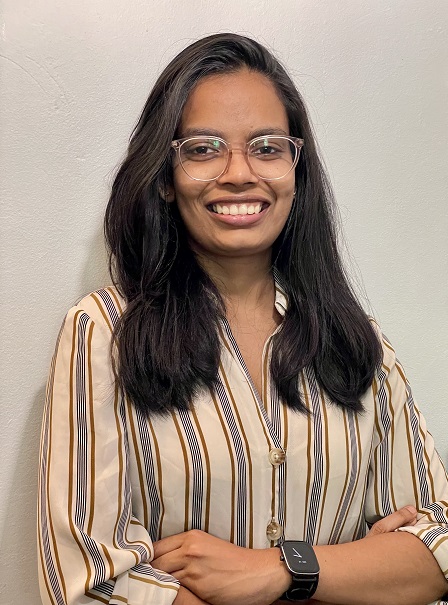}}
Bhagya Chembakottu is a Ph.D. student in Computer Science at McGill University. Her research lies at the intersection of software engineering, security, and human-centered computing, with a focus on analyzing web application security tutorials, static analysis tools, and privacy-aware design practices. Her work combines large-scale empirical analysis with qualitative methods to study how developers and designers engage with security and privacy guidance in practice.
She holds a Master’s degree in Computer and Software Engineering from Polytechnique Montr\'{e}al and a Bachelor’s degree in Computer Science and Engineering from Cochin University of Science and Technology. Contact her at bhagya.chembakottu@mail.mcgill.ca.
\end{IEEEbiography}

\begin{IEEEbiography}{
\includegraphics[width=1.1in]{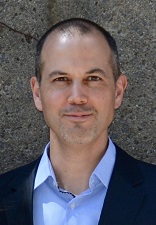}}
Martin P. Robillard is a Professor of Computer Science at McGill University. His research investigates how to facilitate the discovery, acquisition, and effective use of technical, design, and domain knowledge in software development. His work spans software documentation, API usability, program comprehension, and developer productivity.

He served as Program Co-Chair of the 20th ACM SIGSOFT International Symposium on the Foundations of Software Engineering (FSE 2012) and the 39th ACM/IEEE International Conference on Software Engineering (ICSE 2017). He received his Ph.D. and M.Sc. from the University of British Columbia and a B.Eng. from \'{E}cole Polytechnique de Montr\'{e}al. Contact him at robillard@acm.org.
\end{IEEEbiography}

\end{document}